\documentclass[a4paper]{jpconf}
\usepackage{graphicx}
\begin{document}
\title{Nuclear effects in charged-current quasielastic 
neutrino-nucleus scattering}

\author{Maria B. Barbaro}

\address{Universit\`a di Torino, Dipartimento di Fisica Teorica and INFN,
Via P. Giuria 1, 10125 Torino, ITALY}

\ead{barbaro@to.infn.it}

\begin{abstract}

After a short review of the recent developments in studies of
neutrino-nucleus interactions, 
the predictions for double-differential and integrated 
charged current-induced quasielastic cross sections are presented
within two different relativistic approaches:
one is the so-called SuSA method, based on the superscaling behavior exhibited 
by electron scattering data; the other is a microscopic model 
based on relativistic mean field theory, and incorporating final-state 
interactions. 
The role played by the meson-exchange currents
in the two-particle two-hole sector is explored and the results are compared 
with the recent MiniBooNE data.

\end{abstract}

\section{Introduction}

The analysis and interpretation of ongoing and future neutrino oscillation
experiments strongly rely on the nuclear modeling for describing the
interaction of neutrinos and anti-neutrinos with the detector. 
Moreover, neutrino-nucleus scattering has recently become a matter of debate
in connection with the possibility of extracting information on the nucleon
axial mass.
Specifically, the data on muon neutrino charged-current quasielastic (CCQE)
cross sections obtained by the MiniBooNE
collaboration~\cite{AguilarArevalo:2010zc} are substantially underestimated by
the Relativistic Fermi Gas (RFG) prediction.
This has been ascribed either to effects in the elementary
neutrino-nucleon interaction, or to nuclear effects.
The most poorly known ingredient of the single nucleon
cross section is the cutoff parameter $M_A$
employed in the dipole prescription for the axial form factor of the nucleon, 
which can be extracted from $\nu$ and $\overline\nu$ scattering off hydrogen
and deuterium and from charged pion electroproduction.
If $M_A$ is kept as a free parameter in the
RFG calculation, a best fit of the MiniBooNE data yields a value of 
the order of 1.35 GeV/c$^2$, much larger than  the average value 
$M_A\simeq 1.026 \pm 0.021$ GeV/c$^2$ extracted from the (anti)neutrino world 
data~\cite{Bernard:2001rs}. This should be taken more
as an indication of incompleteness of the theoretical description of
the data based upon the RFG, rather than as a true indication for a larger 
axial mass.
Indeed it is well-known from comparisons with electron scattering data
that the RFG model is too crude to account for the nuclear dynamics.
Hence it is crucial to explore more sophisticated nuclear models 
before drawing conclusions on the value of $M_A$. 

Several calculations have been recently performed and applied to neutrino reactions. 
These include, besides the approach that will be presented here, 
models based on 
nuclear spectral functions~\cite{Barbaro:1996vd,Benhar:2005dj,Benhar:2006nr,Benhar:2009wi,Ankowski:2010yh,Benhar:2010nx,Juszczak:2010ve,Ankowski:2011dc}, 
relativistic independent particle models~\cite{Alberico:1997vh,Alberico:1997rm,Alberico:1998qw},
relativistic Green function approaches~\cite{Meucci:2003cv,Meucci:2004ip,Meucci:2006ir,Meucci:2011pi,Meucci:2011vd}, 
models including NN correlations~\cite{Antonov:2006md,Antonov:2007vd,Ivanov:2008ng},
coupled-channel transport models~\cite{Leitner:2006ww,Leitner:2006sp,Buss:2007ar,Leitner:2010kp}, RPA calculations~\cite{Nieves:2004wx,Nieves:2005rq,Valverde:2006zn} and models including multinucleon knock-out~\cite{Martini:2009uj,Martini:2010ex,Nieves:2011pp,Nieves:2011yp}.
The difference between the predictions of the above models can be large due to 
the different treatment of both initial and final state interactions. 
As a general trend, the models based on impulse approximation, where the
neutrino is supposed to scatter off a single nucleon inside the nucleus, 
tend to underestimate the MiniBooNE data, while a sizable increase of the 
cross section is obtained when two-particle-two-hole (2p-2h) mechanisms are 
included in the calculations. Furthermore, a recent calculation performed 
within the relativistic Green function (RGF) framework has shown that at this 
kinematics the results strongly depend on the phenomenological optical 
potential used to describe the final state interaction between the ejected 
nucleon and the residual nucleus~\cite{Meucci:2011vd}.
With an appropriate choice of the optical potential the RGF model 
can reproduce the MiniBooNE data without the need of modifying the axial mass
(see Giusti's contribution to this volume~\cite{Giusti:2011ar}).

The kinematics of the MiniBooNE experiment, where the neutrino flux spans
a wide range of energies reaching values as high as 3 GeV, demands relativity as an essential ingredient. This is illustrated in Fig.~1, where the 
relativistic and non-relativistic Fermi gas results for the 
CCQE double differential cross section of 1 GeV muon neutrinos on $^{12}C$ are 
shown as a function of the outgoing muon momentum and for two values of the
muon scattering angle. The relativistic effects, which affect both the 
kinematics and the dynamics of the problem, have been shown to be relevant 
even at moderate momentum and energy transfers~\cite{Amaro:1998ta,Amaro:2002mj}.
\begin{figure}[h]
\begin{minipage}{38pc}
\includegraphics[height=10pc]{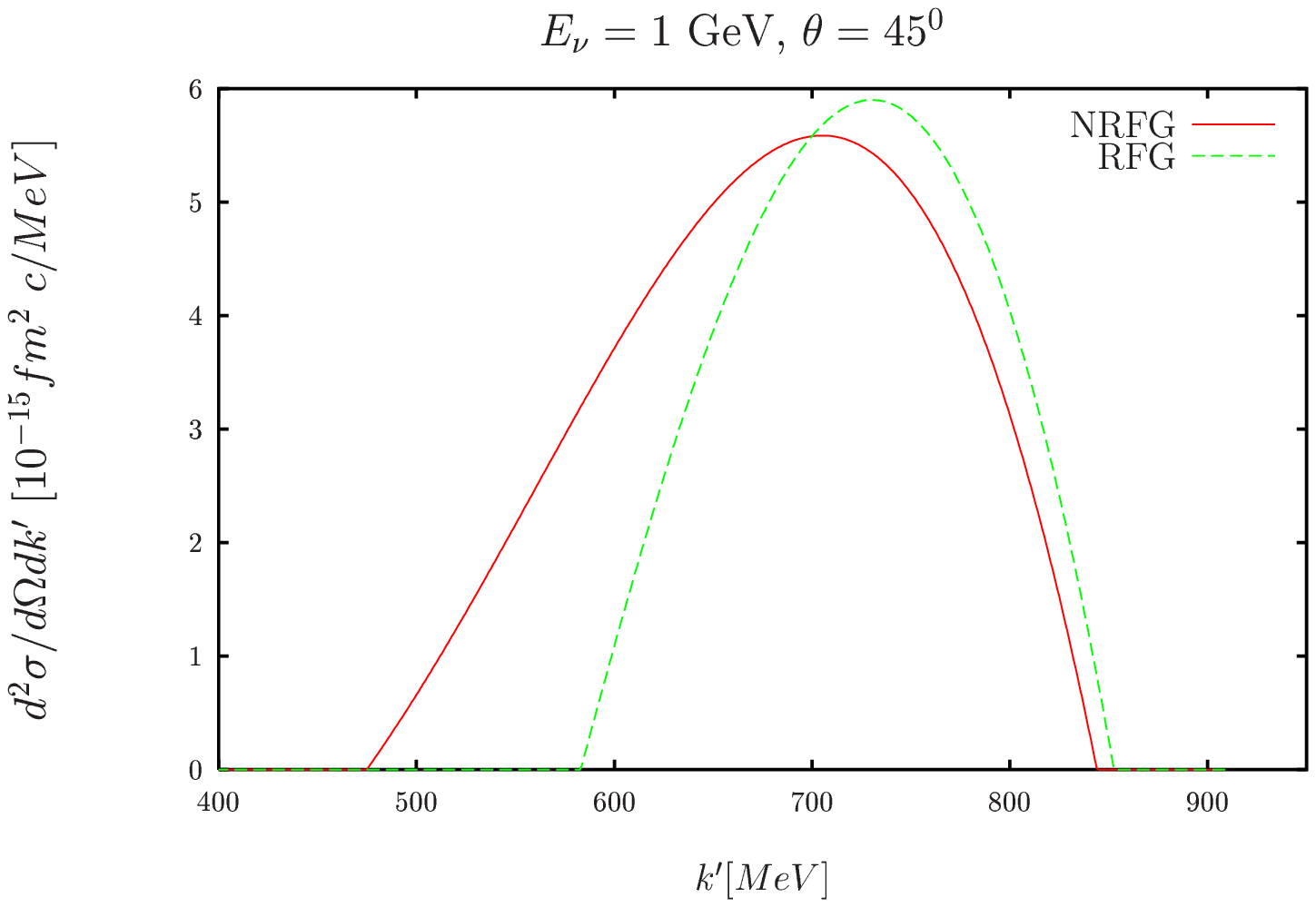}
\hspace{3pc}
\includegraphics[height=10pc]{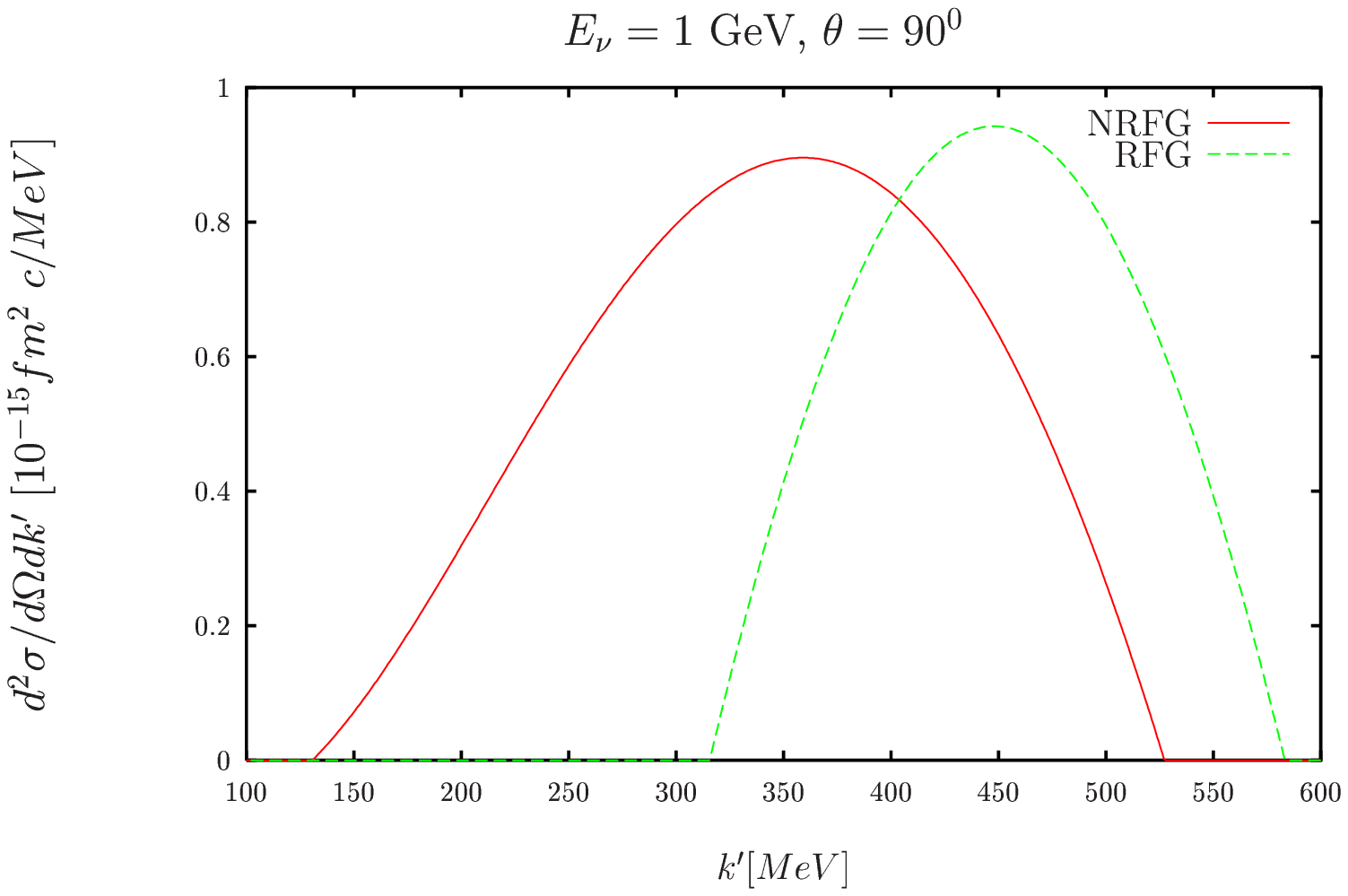}
\caption{\label{fig:nr} $\nu_\mu$CCQE 
double differential cross sections on $^{12}C$ 
displayed versus the outgoing muon momentum for non-relativistic (NRFG) and relativistic (RFG) Fermi gas.}
\end{minipage}\hspace{2pc}%
\end{figure}
Hence in our approach we try to retain as much as possible 
the relativistic aspects of the problems.

In spite of its simplicity, the RFG has the merit of incorporating 
an exact relativistic treatment, fulfilling the fundamental properties of
Lorentz covariance and gauge invariance. However, it badly fails to reproduce the electron
scattering data, in particular when it is compared with the Rosenbluth-separated
longitudinal and transverse responses. 
Comparison with electron scattering data must be
a guiding principle in selecting reliable models for neutrino reactions. 
A strong constraint in this connection is represented
by the ``superscaling'' analysis of the world inclusive $(e,e')$ 
data: in Refs.~\cite{Day:1990mf,Jourdan:1996ut,Donnelly:1998xg,Donnelly:1999sw,
Maieron:2001it} it has been proved that, for sufficiently large momentum
transfers, the reduced cross section (namely the double differential
cross section divided by the appropriate single nucleon factors), 
when represented versus the scaling variable $\psi$~\cite{Alberico:1988bv}, 
is largely independent of the momentum transfer (first-kind scaling) and of 
the nuclear target (second-kind scaling). The simultaneous occurrence of the
two kinds of scaling is called susperscaling. 
Moreover, from the experimental longitudinal response a
phenomenological quasielastic scaling function has been extracted that shows a
clear asymmetry with respect to the quasielastic peak (QEP) 
with a long tail extended to
positive values of the scaling variable, i.e., larger energy transfers.
On the contrary the RFG model, as well as most models based on impulse
approximation, give a symmetric superscaling function with a maximum value 
20-30\% higher than the data~\cite{Barbaro:2006me}. 

In this contribution, after recalling the basic formalism for CCQE reactions
and their connection with electron scattering,
we shall illustrate two models which provide good agreement with the 
above properties of electron scattering data: one of them, the relativistic
mean field (RMF) model, 
comes from microscopic many-body theory, the other, the 
superscaling approximation (SuSA) model, is 
extracted from $(e,e')$ phenomenology. 
We shall then include the contribution of 2p-2h excitations in the SuSA
model and finally compare our results with the MiniBooNE double differential,
single differential and total cross sections.
Most of the results which will be presented are contained in 
Refs.~\cite{Amaro:2010sd} and \cite{Amaro:2011qb}.

\section{Formalism}

Charged current quasielastic  
muonic neutrino scattering $(\nu_\mu,\mu^{-})$ off nuclei 
is very closely related to quasielastic inclusive electron scattering $(e,e')$.
However two major differences occur between the two processes:
\begin{enumerate}
\item in the former case the probe interacts with the nucleus via the weak
force, in the latter the interaction is (dominantly) electromagnetic. 
While the weak vector
current is related to the electromagnetic one via the CVC theorem, 
the axial current gives rise to a more complex structure of the cross sections,
with new response functions which cannot be related to the electromagnetic ones.
As a consequence, while in electron scattering the double-differential cross 
section can be expressed in terms of two response functions, longitudinal and
transverse with respect to three-momentum carried by the virtual photon, for 
the CCQE process it can be written according to a Rosenbluth-like decomposition 
as~\cite{Amaro:2004bs}
\begin{equation}
\left[ \frac{d^2\sigma}{dT_\mu d\cos\theta} \right]_{E_\nu} =
\sigma_0 \left[ {\hat V}_{L} R_L
+ {\hat V}_T R_T + {\hat V}_{T^\prime} R_{T^\prime} \right] ,
\label{eq:d2s}
\end{equation}
where $T_\mu$ and $\theta$ are the muon kinetic energy and scattering angle,
$E_\nu$ is the incident neutrino energy, $\sigma_0$ is the elementary 
cross section, ${\hat V}_i$ are kinematic 
factors and $R_i$ are the nuclear response functions, 
the indices $L,T,T^\prime$ referring to longitudinal, transverse,
transverse-axial, components of the nuclear current, respectively.
The expression (\ref{eq:d2s}) is formally analogous to the inclusive electron 
scattering case, but: {\it i)} the ``longitudinal'' response $R_L$
takes contributions from the charge (0) and longitudinal (3) components of the 
nuclear weak current, which, at variance with the electromagnetic case, 
are not related to each other by current conservation, {\it ii)} $R_L$ and $R_T$
have both ``VV'' and ``AA'' components (stemming from the product of two vector
or axial currents, respectively),
{\it iii)} a new
response, $R_{T^\prime}$, arises from the interference between the axial and vector
parts of the weak nuclear current.

In Fig.~2 we show the separate contributions of the three responses in 
(\ref{eq:d2s}), evaluated
in the RFG model for the $^{12}C$ target nucleus, for two different values of
the scattering angle. 
It can be seen that in the forward direction the transverse response dominates 
over the longitudinal and transverse-axial ones, whereas at higher angles the
$L$-component becomes negligible and the $T^\prime$ and $T$ responses are almost
equal. This cancellation has important consequences on antineutrino-nucleus
scattering, where the response $R_{T^\prime}$ has opposite sign.

\begin{figure}[h]
\begin{minipage}{38pc}
\includegraphics[width=16pc]{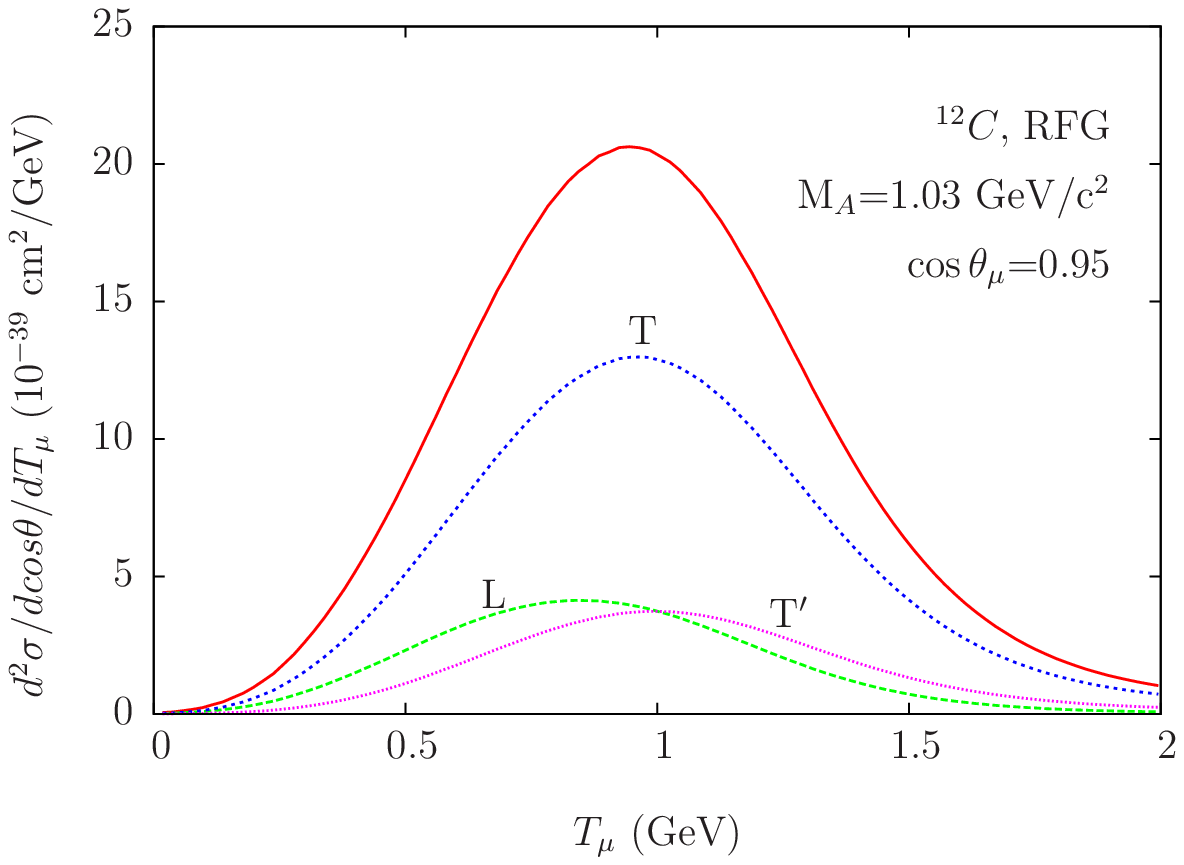}
\hspace{3pc}
\includegraphics[width=16pc]{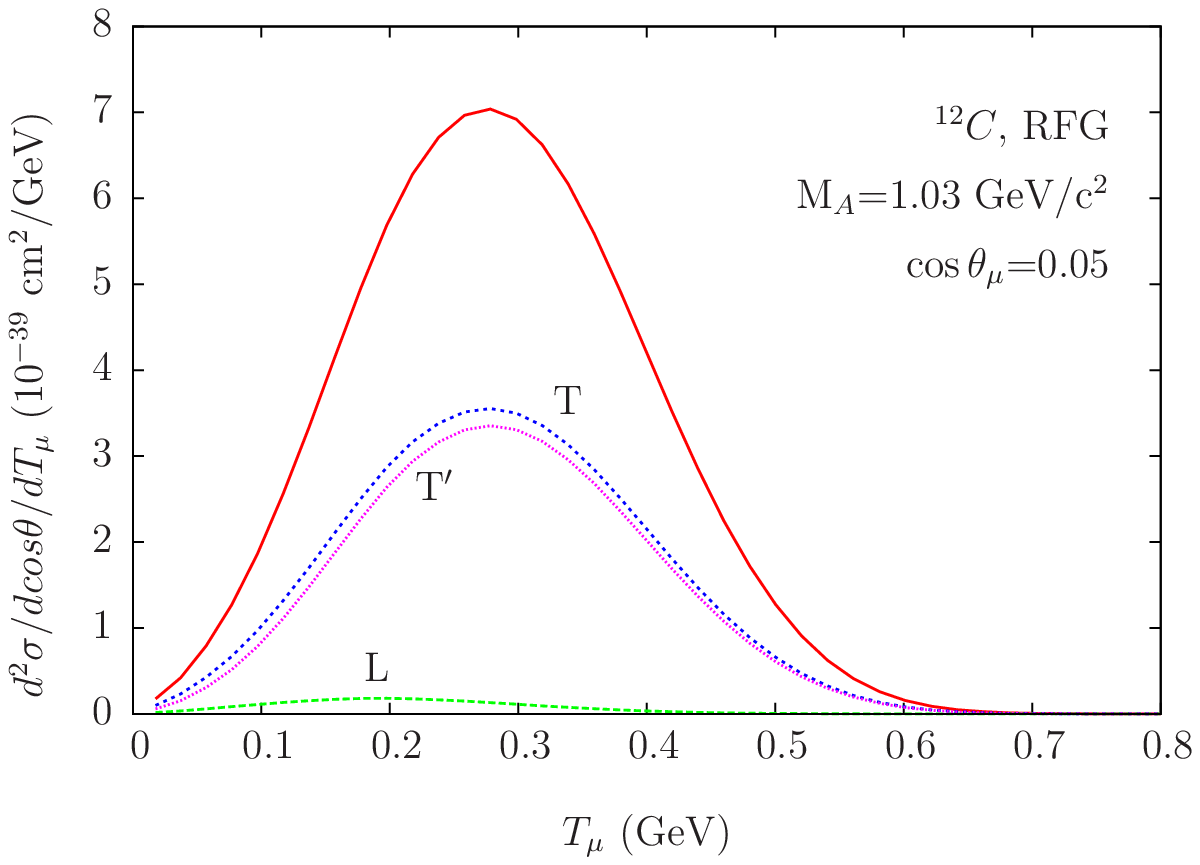}
\caption{\label{fig:sep} Separate contributions of the RFG longitudinal $(L)$,
transverse $(T)$ and axial-vector interference $(T^\prime)$ responses to the 
double differential $\nu_\mu$CCQE cross sections displayed versus the muon kinetic
energy at two different angles. The neutrino energy is averaged over the 
MiniBooNE flux and the axial mass parameter has the standard value.
}
\end{minipage}\hspace{2pc}%
\end{figure}

\item In $(e,e')$ experiments the energy of the electron is well-known, and therefore the detection of the outgoing electron univoquely determines
the energy and momentum transferred to the nucleus. 
In neutrino experiments the neutrino energy is not known, but distributed over a range of
values (for MiniBooNE from 0 to 3 GeV with an average value 
of about 0.8 GeV). 
The cross section must then be evaluated
as an average over the experimental flux $\Phi(E_\nu)$ 
\begin{equation}
\frac{d^2\sigma}{dT_\mu d\cos\theta}
= \frac{1}{\Phi_{tot}} \int
\left[ \frac{d^2\sigma}{dT_\mu d\cos\theta} \right]_{E_\nu}
\Phi(E_\nu) dE_\nu \ ,
\label{eq:fluxint}
\end{equation}
which may require to
account for effects not included in models devised for quasi-free scattering. 
This is, for instance, the situation at the most forward scattering angles, 
where a significant contribution in the cross section comes from
very low-lying excitations in nuclei~\cite{Amaro:2010sd}, as illustrated in Fig.~3:
here the double differential cross section is evaluated in the SuSA model 
(see later) at the MiniBooNE kinematics and the lowest angular bin and compared
with the result obtained by excluding the energy transfers lower than 50 MeV
from the flux-integral (\ref{eq:fluxint}). 
\begin{figure}[h]
\begin{minipage}{17pc}
\includegraphics[width=17pc]{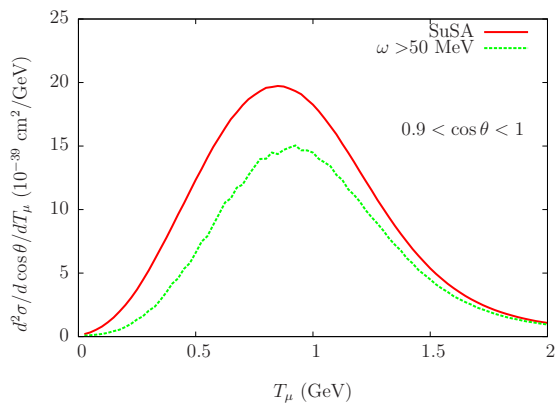}
\end{minipage}\hspace{3pc}%
\begin{minipage}{17pc}
\caption{\label{fig:cut}(Color online) Solid lines (red online): 
flux-integrated $\nu_\mu$CCQE cross sections on $^{12}C$ calculated in the SuSA 
model for a specific bin of scattering angle. Dashed lines (green online): 
a lower cut $\omega=50$ MeV is set in the integral over the neutrino flux.}
\end{minipage}%
\end{figure}
It appears that at these angles 30-40\% of the cross section corresponds
to very low energy transfers, where collective effects dominate.
Moreover, processes involving meson exchange currents (MEC), which can excite
both one-particle-one-hole (1p1h) and two-particle-two-hole (2p-2h) states
via the exchange of a virtual meson, 
should also be taken into account, since they lead to final states where no
pions are present, classified as ``quasielastic'' in the MiniBooNE
experiment.
\end{enumerate}

\section{Models}

In this Section we briefly outline the main ingredients of the RMF and SuSA 
model and
we illustrate our calculation of the contribution of 2p-2h meson 
exchange currents.

\subsection{RMF}

In the RMF model 
a fully relativistic description of both the kinematics and the 
dynamics of the process is incorporated.

Details on the RMF model applied to inclusive QE electron and CCQE neutrino
reactions can be found in Refs.~\cite{Caballero:2006wi,Caballero:2007tz,Caballero:2005sj,Amaro:2006tf,Maieron:2003df,Amaro:2006if}.
Here we simply recall that the weak response functions are given by 
taking the appropriate components of the weak hadronic tensor, 
constructed from the single-nucleon current matrix elements
\begin{equation}
\langle J_W^\mu\rangle = \int d{\bf r}\overline{\phi}_F({\bf r})\hat{J}_W^\mu({\bf r})\phi_B({\bf r})
\, ,
\end{equation}
where $\phi_B$ and $\phi_F$ are relativistic bound-state and
scattering wave functions, respectively, and $\hat{J}_W^\mu$ is
the relativistic one-body current operator modeling the coupling
between the virtual $W$-boson and a nucleon. The bound
nucleon states are described as self-consistent Dirac-Hartree
solutions, derived by using a Lagrangian containing $\sigma$, $\omega$ and 
$\rho$
mesons~\cite{Horowitz,Serot,Sharma:1993it}. The outgoing nucleon
wave function is computed by using the same relativistic mean
field (scalar and vector energy-independent potentials) employed in the 
initial state and incorporates the final state interactions (FSI)
between the ejected proton and the residual nucleus.

The RMF model successfully 
reproduces the scaling behaviour of  inclusive QE $(e,e')$ processes 
and, more importantly, it gives rise to a
superscaling function with a significant asymmetry, namely, in
complete accord with data~\cite{Caballero:2006wi,Caballero:2007tz}.
This is a peculiar property associated to the consistent treatment of initial
and final state interactions. It has been shown 
in Refs.~\cite{Caballero:2006wi,Caballero:2007tz} that other versions of the RMF 
model, which deal with the FSI through a real relativistic optical potential, are not
capable of reproducing the asymmetry of the scaling function.

Moreover, contrary to most other models based on impulse approximation, 
where scaling of the 
zeroth kind - namely the equality of the longitudinal and transverse scaling functions -  occurs, the RMF model provides $L$ and $T$ scaling
functions which differ by typically $20\%$, the T one being larger.
This agrees with the analysis \cite{Donnelly:1999sw} of
the existing $L/T$ separated data, which has shown that, after
removing inelastic contributions and two-particle-emission effects,
the purely nucleonic transverse scaling function
is significantly larger than the longitudinal one.

\subsection{SuSA}

The SuSA model is based on the phenomenological superscaling function
extracted from the world data on quasielastic electron 
scattering~\cite{Jourdan:1996ut}. The model has been extended to the 
$\Delta$-resonance region in Ref.~\cite{Amaro:2004bs} and to neutral current
scattering in Ref.~\cite{Amaro:2006pr}, but here we restrict our attention 
to the quasielastic charged current case.

Assuming the scaling function $f$ extracted from $(e,e')$ data 
to be universal, i.e., valid for electromagnetic and weak interactions,
in~\cite{Amaro:2004bs,Amaro:2005dn} CCQE neutrino-nucleus cross
sections have been evaluated by multiplying $f$ by the corresponding elementary
weak cross section. 
Thus in the SuSA approach all
the nuclear responses in (\ref{eq:d2s}) are expressed as follows
\begin{equation}
R_K = N \frac{2 E_F}{k_F q} U_K(q,\omega) f(\psi) \ \ , \ \ \ (K=L,T,T^\prime)
\end{equation}
where $U_K$ are the elementary lepton-nucleon responses, $E_F$ and $k_F$ are
the Fermi energy and momentum, $N$ is the number of nucleons (neutrons
in the $\nu_\mu$CCQE case) and $f(\psi)$ in the universal superscaling function,
depending only on the scaling variable $\psi$~\cite{Alberico:1988bv}.

The SuSA approach 
provides nuclear-model-independent neutrino-nucleus cross sections
and reproduces the longitudinal electron data by construction.
However, its reliability rests on some basic assumptions.

First, it assumes that the scaling function - extracted from {\it longitudinal} 
$(e,e')$ data - is appropriate 
for all of the weak responses involved in neutrino
scattering (charge-charge, charge-longitudinal,
longitudinal-longitudinal, transverse and axial), and is independent
of the vector or axial nature of the nuclear current entering the
hadronic tensor. In particular 
it assumes the equality of the longitudinal and transverse
scaling functions (scaling of the zeroth kind), which, as mentioned before, 
has been shown to be violated both by experiment and by some microscopic models,
for example relativistic mean field theory.

Second, the charged-current neutrino responses are purely isovector, whereas the
electromagnetic ones contain both isoscalar and isovector
components and the former involve axial-vector as well as vector responses. 
One then has to invoke a further kind of scaling, namely
the independence of the scaling function of the choice of isospin
channel --- so-called scaling of the third kind. 

Finally, the SuSA approach neglects violations of scaling of first and second kinds. These are known to be important at energies above the QE peak and to reside mainly in the transverse channel, being associated to effects which go 
beyond the
impulse approximation: inelastic scattering, meson-exchange currents
and the associated correlations needed to conserve the vector current.
The inclusion of these contributions in the SuSA model is discussed 
in the next paragraph.

\subsection{2p-2h MEC}

\begin{figure}[h]
\includegraphics[width=25pc]{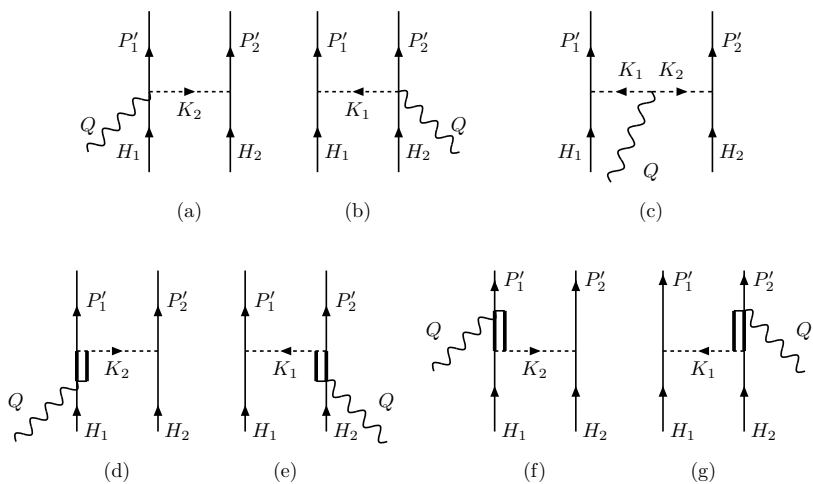}\hspace{2pc}%
\begin{minipage}{11pc}
\vspace{-5pc}%
\caption{\label{fig:mec} Two-body meson-exchange currents. 
(a) and (b): ``contact'', or ``seagull'' diagram;  
(c): ``pion-in-flight'' diagram;
(d)-(g): ``$\Delta$-MEC'' diagram.}
\end{minipage}
\end{figure}

Meson exchange currents are two-body currents carried by a virtual meson
which is exchanged between two nucleons in the nucleus. They are represented by
the diagrams in Fig.~\ref{fig:mec}, where the external lines correspond to 
the virtual boson ($\gamma$ or $W$) and the dashed lines to the exchanged 
meson: in our approach we only consider the pion, which is believed to give the
dominant contribution in the quasielastic regime. The thick lines in diagrams
(d)-(g) represent the propagation of a $\Delta$-resonance.
The explicit relativistic expressions for the current matrix elements can be found, e.g., in Ref.~\cite{Amaro:2010iu}.

Being two-body currents, the MEC can excite both one-particle one-hole 
(1p-1h) and two-particle two-hole (2p-2h) states.
In the 1p-1h sector, 
MEC studies of electromagnetic $(e,e^\prime)$ 
process have been performed for low-to-intermediate momentum transfers
(see, {\it e.g.}, \cite{Alberico:1989aja,Amaro:2002mj,Amaro:2003yd,Amaro:2009dd}),
showing a small reduction of the total response at the quasielastic peak,
mainly due to diagrams involving the electroexcitation of the
$\Delta$ resonance. 

However in a perturbative scheme where all the diagrams containing one and only
one pionic line are retained, the MEC are not the only diagrams arising,
but pionic correlation contributions, where the virtual boson is attached to 
one of the two interacting nucleons, should also be considered.
These are represented by the same diagrams as in Fig.~\ref{fig:mec}(d)-(g), where now the thick lines are nucleon propagators. Only when all the diagrams
are taken into account gauge invariance is fulfilled and the full two-body
current is conserved.
Correlation diagrams have been shown to roughly compensate the pure MEC
contribution~\cite{Alberico:1989aja,Amaro:2002mj,Amaro:2003yd,Amaro:2009dd}, 
so that in first approximation we can neglect the 1p-1h sector 
and restrict our attention to 2p-2h final states.

The contribution to the inclusive electron scattering cross section 
arising from two-nucleon emission via meson exchange current interactions 
was first calculated in the Fermi gas model in 
Refs.~\cite{Donnelly:1978xa,VanOrden:1980tg}, where sizable effects 
were found at large energy transfers.
In these references a non-relativistic reduction of the
currents was performed, while fully relativistic calculations have been
developed more recently in Refs.~\cite{Dekker:1994yc,De Pace:2003xu,Amaro:2010iu}. It
has been found that the MEC give a significant positive contribution
which leads to a partial filling of the ``dip'' between the quasielastic peak 
and the analogous peak associated with the excitation of the $\Delta$ resonance.
Moreover, the MEC
have been shown to break scaling of both first and second
kinds~\cite{De Pace:2004cr}.

Here we use the fully relativistic model of \cite{De Pace:2003xu}, where all the MEC many-body diagrams containing two pionic
lines that contribute to the electromagnetic 2p-2h transverse response
were taken into account. 
Similar results for the 2p-2h MEC were obtained in 
Ref.~\cite{Amaro:2010iu}, where the correlation diagrams were also included.

In order to apply the model to neutrino scattering,
we observe that in lowest order the 2p2h sector is not directly reachable 
for the axial-vector matrix elements. Hence the MEC affect only
the transverse polar vector response, $R_T^{VV}$.  Note that, at variance with the 1p-1h sector, where the contribution of the
MEC diagrams originates from the interference between 1-body and 2-body 
amplitudes and has therefore no definite sign (in fact it turns out to be 
negative due to the dominance of the diagrams involving the $\Delta$), 
the 2p-2h contribution of MEC to the nuclear responses is the square of an 
amplitude, hence it is positive by definition. 
Therefore the net effect of 2p-2h MEC to neutrino scattering is to increase
the transverse vector response function, as will be illustrated in the next section.

\section{Results}

In this Section we present the predictions of the above models and their 
comparison with the MiniBooNE data. More results can be found in 
Refs.~\cite{Amaro:2010sd} and \cite{Amaro:2011qb}.

\begin{figure}[h]
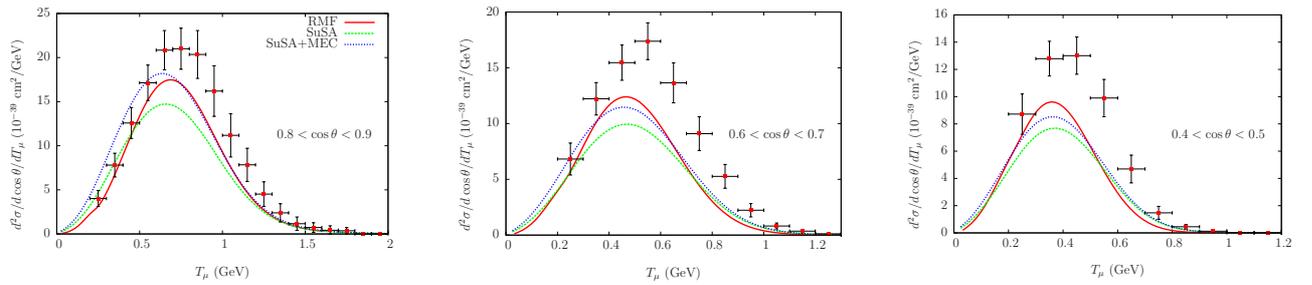

\begin{minipage}{12pc}
\includegraphics[width=12pc]{RMF_c085.epsi}
\end{minipage}\hspace{2pc}%
\begin{minipage}{12pc}
\includegraphics[width=12pc]{RMF_c065.epsi}
\end{minipage}\hspace{2pc}%
\begin{minipage}{12pc}
\includegraphics[width=12pc]{RMF_c045.epsi}
\end{minipage} 
\caption{\label{fig:cos} Flux-integrated double-differential cross
section per target nucleon for the $\nu_\mu$ CCQE process on
$^{12}$C evaluated in the RMF (red) model and in the SuSA approach 
with (blue line) and without (green line) the contribution of MEC and 
displayed versus the muon kinetic energy $T_\mu$ for three specific bins
of the scattering angle. 
The data are from MiniBooNE~\cite{AguilarArevalo:2010zc}.}
\end{figure}

\begin{figure}[h]
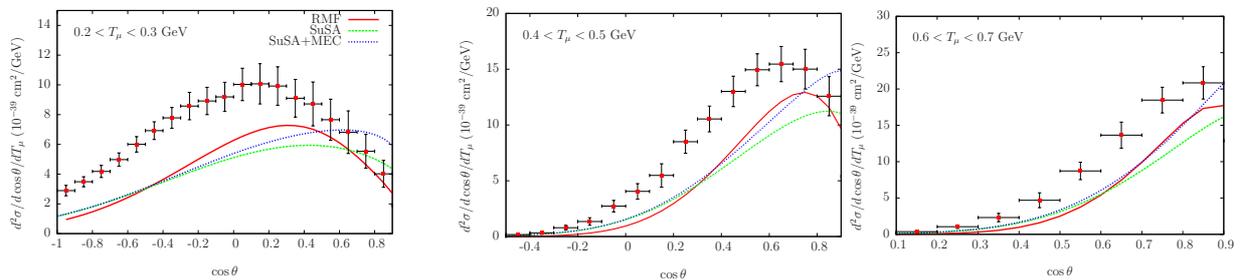

\begin{minipage}{12pc}
\includegraphics[width=12pc]{RMF_t025.epsi}
\end{minipage}\hspace{2pc}%
\begin{minipage}{12pc}
\includegraphics[width=12pc]{RMF_t045.epsi}
\end{minipage} 
\begin{minipage}{12pc}
\includegraphics[width=12pc]{RMF_t065.epsi}
\end{minipage} 
\caption{\label{fig:tmu} Flux-integrated double-differential cross
section per target nucleon for the $\nu_\mu$ CCQE process on
$^{12}$C evaluated in the RMF (red) model and in the SuSA approach 
with (blue line) and without (green line) the contribution of MEC and 
displayed versus the muon scattering angle for three bins of the
muon kinetic energy $T_\mu$.
The data are from MiniBooNE~\cite{AguilarArevalo:2010zc}.
}
\end{figure}

In Figs.~\ref{fig:cos} and \ref{fig:tmu} the flux-integrated 
double-differential cross
section per target nucleon for the $\nu_\mu$CCQE process on 
$^{12}$C is evaluated for the three nuclear models above described: 
the RMF model and
the SuSA approach with and without the contribution of 2p-2h MEC.
In Fig.~\ref{fig:cos} the cross sections are displayed versus the muon
kinetic energy $T_\mu$ at fixed scattering energy $\theta$, in Fig.~\ref{fig:tmu}
they are displayed versus $\cos\theta$ at fixed $T_\mu$.
 
It appears that the SuSA predictions systematically underestimate the 
experimental cross section, the discrepancy being larger at high scattering 
angle and low muon kinetic energy.
The inclusion of 2p-2h MEC tends to improve the agreement with the data at low angles,
but it is not sufficient to account for the discrepancy at higher angles.
The RMF calculation, which, as already mentioned, incorporates violations of 
scaling of the zeroth kind with a substantial enhancement of the vector
transverse response, yields cross sections which are in general larger than 
the SuSA ones. 
In particular, in the region close to the peak in the cross section,
the RMF result becomes larger than the one obtained with SuSA+MEC. Furthermore,
the RMF does better than SuSA in fitting the shape of the experimental curves
versus both the scattering angle and the muon energy: 
this is partly due to the fact that the RMF is better describing the 
low-energy excitation region whereas the SuSA model has no predictive power at very low angles, where the cross section is dominated by low excitation energies and the superscaling ideas are not supposed to apply.
                                                       
Concerning the SuSA+MEC results, a possible explanation of the theory/data 
disagreement is the fact that, 
as already mentioned, a fully consistent treatment of two-body currents 
should take into account not only the genuine MEC contributions, but also
the correlation diagrams that are necessary in order to preserve the gauge invariance of the theory. This, however, is not an easy task because
in an infinite system like the RFG the correlation diagrams give rise to
divergences which need to be regularized~\cite{Amaro:2010iu}. The divergences 
arise from a double pole in some of the diagrams, associated to the presence
of on-shell nucleon propagators. Different prescriptions have been used in the
literature in order to overcome this problem~\cite{Alberico:1983zg,Alberico:1990fc,Gil:1997bm,Amaro:2010iu},
leading to a substantial model-dependence of the results.
In particular in Ref.~\cite{Amaro:2010iu} the divergence has been cured by introducing a 
parameter $\epsilon$ which accounts for the finite size of the nucleus (and therefore
the finite time of propagation of a nucleon inside the nucleus) and the $\epsilon$-dependence
of the contribution of correlation diagrams has been explored. The study has shown that
for reasonable values of the parameter the correlations add to the pure MEC in the 
high-energy tail and are roughly of the same order of magnitude, but now contributing
to both the longitudinal and the transverse channels. The inclusion of these terms in neutrino 
reactions is in progress~\cite{Amaro:wip} and is expected to give a further enhancement of the
cross sections.

The single 
differential cross sections with respect to the muon kinetic
energy and scattering angle, respectively, are presented
in Figs.~\ref{fig:csvst} and \ref{fig:csvscos}, where the relativistic Fermi 
gas result is also shown for comparison:
again it appears that the RMF gives slightly higher cross sections
than SuSA, due to the $L/T$ unbalance, but both models still underestimate 
the data for
most kinematics. The inclusion of 2p-2h excitations leads to a good agreement 
with the data at high $T_\mu$, but strength is still missing at the lower muon 
kinetic energies (namely higher energy transfers) and higher angles.

\begin{figure}[h]
\begin{minipage}{17pc}
\includegraphics[width=17pc]{dsdcos.epsi}
\caption{\label{fig:csvst}(Color online) 
Flux-averaged 
$\nu_\mu$CCQE cross section on $^{12}C$ 
integrated over the scattering angle and
displayed versus the muon kinetic energy.
The data are from MiniBooNE~\cite{AguilarArevalo:2010zc}.
}
\end{minipage}\hspace{2pc}%
\begin{minipage}{17pc}
\includegraphics[width=17pc]{dsdt.epsi}
\caption{\label{fig:csvscos}(Color online) 
Flux-averaged 
$\nu_\mu$CCQE cross section on $^{12}C$ 
integrated over the muon kinetic energy
and displayed versus the scattering angle.
The data are from MiniBooNE~\cite{AguilarArevalo:2010zc}.
}
\end{minipage} 
\end{figure}

\begin{figure}[h]
\begin{minipage}{17pc}
\includegraphics[width=17pc]{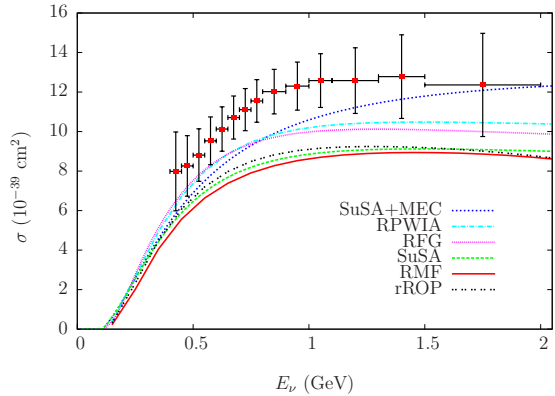}\hspace{2pc}%
\end{minipage}\hspace{3pc}%
\begin{minipage}{17pc}\caption{\label{fig:csvsenu}(Color online)
Total CCQE cross section 
per neutron versus the neutrino energy. The curves corresponding to different
nuclear models are compared with the flux unfolded
MiniBooNE data~\cite{AguilarArevalo:2010zc}.}
\end{minipage}
\end{figure}

Finally, in Fig.~\ref{fig:csvsenu} the total 
(namely integrated over over all muon scattering angles and energies)
CCQE cross section per neutron is displayed versus the neutrino energy 
and compared with the experimental
flux-unfolded data. Besides the models above discussed, we show
for comparison also the results of the relativistic mean field model 
when the final state interactions are ignored (denoted as RPWIA - relativistic
plane wave impulse approximation)
or described through a real optical potential (denoted as rROP).
Note that the discrepancies between the various models, observed in
Figs.~\ref{fig:cos} and \ref{fig:tmu},
tend to be washed out by the integration, yielding very similar
results for the models that include FSI (SuSA, RMF and rROP), all of
them giving a lower total cross section than the models without FSI
(RFG and RPWIA). On the other hand the SuSA+MEC curve, while being
closer to the data at high neutrino energies, has a somewhat
different shape with respect to the other models, in qualitative
agreement with the relativistic calculation of \cite{Nieves:2011pp}.
It should be noted, however, that the result is affected by an
uncertainty of about 5\% associated with the treatment of the
2p-2h contribution at low momentum transfers and that pionic correlations
are not included.

\section{Conclusions}

Two different relativistic models, one (SuSA)
phenomenological and the other (RMF) microscopic, have been applied to the study
of charged-current quasielastic neutrino scattering and 
the impact of 2p-2h meson exchange currents on the cross sections has been
investigated.
The results can be summarized as follows:

\begin{enumerate}

\item 
Both the SuSA and the RMF models, in contrast with the
relativistic Fermi gas, are fitting with good accuracy the longitudinal 
quasielastic electron scattering response at intermediate to high energy and 
momentum transfer. 

The SuSA and RMF models give very similar results for the integrated neutrino
cross section and both substantially under-predict the MiniBooNE experimental
data. However the comparison with the double differential experimental 
cross section reveals some differences between the two models, which are washed 
out by the integration. Indeed the RMF, although being lower than the data, 
reproduces better the slopes of the cross section versus the muon energy and 
scattering angle. This is essentially due to the enhancement of the transverse
response, which arises from the self-consistent mean field approach of RMF
(in particular from the consistent treatment of initial and final state 
interactions) and is absent in the superscaling approach.

\item

In relativistic
or semi-relativistic models final state interactions have been shown to play 
an essential role for reproducing the shape and size of the electromagnetic 
response~\cite{Maieron:2003df,Caballero:2006wi,Amaro:2006if} 
and cannot be neglected, in our scheme, in the study of neutrino interactions.
The effect of final state interactions in the SuSA and RMF models is to 
lower the cross section, giving 
a discrepancy with the data larger than the RFG. 

\item

In the transverse channel, the analysis of $(e,e')$ data points to the
importance of meson-exchange currents which, through the excitation of 
two-particle-two-hole states, are partially responsible of filling 
the ``dip'' region between the QE and $\Delta$ peaks. The 2p-2h MEC can be
even more relevant in the CCQE  process, where ``quasielastic'' implies simply that
no pions are present in the final state but, due to the large energy region 
spanned by the neutrino flux, processes involving the exchange of virtual 
pions can give a sizable contribution.
In fact the inclusion of 2p-2h MEC
contributions yields larger cross sections and accordingly better
agreement with the data, although the theoretical curves still lie 
below the data at high angles and low muon energy.
It should be stressed, however, that the present calculation, though exact and
fully relativistic, is incomplete. In order to preserve gauge invariance
the full two-body current,
including not only the MEC but also the corresponding correlation diagrams, must
be included. These have recently 
been shown to yield a sizable contribution at high 
energies in $(e,e')$ scattering~\cite{Amaro:2010iu} and are likely to 
improve the agreement of our models with the MiniBooNE data.

\item

In all our calculations the standard value $M_A=$1.03 GeV/c$^2$ has been used. 
It has been suggested that a larger value of the axial mass (1.35 GeV/c$^2$)
would eliminate the disagreement with the data. However the fit was done
using a RFG analysis, and more sophisticated nuclear models must be explored
before drawing conclusions on the actual value of the axial mass. 
For instance in Ref.~\cite{Bodek:2011ps} it is shown that the MiniBooNE
data can as well be fitted by effectively incorporating some nuclear effects 
in the magnetic form factor of the bound nucleon, without changing the axial 
mass. 

Although our scope here is not to extract a value for the axial mass of
the nucleon, but rather to understand which nuclear effects are 
effectively accounted for by a large axial cutoff parameter, let us
mention that a best fit of the RMF and SuSA results to the MiniBooNE
experimental cross section gives an effective axial mass
$M_A^{\rm eff}\simeq$ 1.5 GeV/c$^2$
and values in the range $1.35<M_A^{\rm eff}<1.65$ GeV/c$^2$ yield
results compatible with the MiniBooNE data within the experimental errors.
A similar analysis in the model including the 2p-2h contribution will be 
possible only when the above mentioned correlation diagrams will be 
consistently evaluated~\cite{Amaro:wip}.

\end{enumerate}

\section*{Acknowledgments}
{I would like to thank
J.E. Amaro, J.A. Caballero, T.W. Donnelly, J.M. Udias and C.W. Williamson for 
the fruitful collaboration which lead to the results reported in this 
contribution.
}


\section*{References}
\begin{thebibliography}{9}
% 1
\bibitem{AguilarArevalo:2010zc}
  A.~A.~Aguilar-Arevalo {\it et al.}  [MiniBooNE Collaboration],
  %``First Measurement of the Muon Neutrino Charged Current Quasielastic Double
  %Differential Cross Section,''
  Phys.\ Rev.\ D {\bf 81}, 092005 (2010).
  %arXiv:1002.2680 [hep-ex].
  %%CITATION = ARXIV:1002.2680;%%
% 2
\bibitem{Bernard:2001rs}
  V.~Bernard, L.~Elouadrhiri and U.~G.~Meissner,
  %``Axial structure of the nucleon: Topical Review,''
  J.\ Phys.\ G {\bf 28}, R1 (2002).
%  [arXiv:hep-ph/0107088].
  %%CITATION = JPHGB,G28,R1;%%
% 3
\bibitem{Barbaro:1996vd}
  M.~B.~Barbaro, A.~De Pace, T.~W.~Donnelly, A.~Molinari and M.~J.~Musolf,
  %``Probing nucleon strangeness with neutrinos: Nuclear model dependences,''
  Phys.\ Rev.\  C {\bf 54}, 1954 (1996).
  %%CITATION = PHRVA,C54,1954;%%
% 4
\bibitem{Benhar:2005dj}
  O.~Benhar, N.~Farina, H.~Nakamura, M.~Sakuda and R.~Seki,
  %``Electron- and neutrino-nucleus scattering in the impulse approximation
  %regime,''
  Phys.\ Rev.\  D {\bf 72}, 053005 (2005).
% 5
\bibitem{Benhar:2006nr}
  O.~Benhar and D.~Meloni,
  %``Total neutrino and antineutrino nuclear cross-sections around 1-GeV,''
  Nucl.\ Phys.\  A {\bf 789}, 379 (2007).
 % [arXiv:hep-ph/0610403].
  %%CITATION = NUPHA,A789,379;%%
% 6
\bibitem{Benhar:2009wi}
  O.~Benhar and D.~Meloni,
  %``Impact of nuclear effects on the determination of the nucleon axial mass,''
  Phys.\ Rev.\  D {\bf 80}, 073003 (2009).
   %%CITATION = PHRVA,D80,073003;%%
% 7
\bibitem{Ankowski:2010yh}
  A.~M.~Ankowski, O.~Benhar and N.~Farina,
  %``Analysis of the Q^2-dependence of charged-current quasielastic processes in
  %neutrino-nucleus interactions,''
  Phys.\ Rev.\  D {\bf 82}, 013002 (2010).
  %%CITATION = PHRVA,D82,013002;%%
% 8
\bibitem{Benhar:2010nx}
  O.~Benhar, P.~Coletti and D.~Meloni,
  %``Electroweak nuclear response in quasi-elastic regime,''
  Phys.\ Rev.\ Lett.\  {\bf 105}, 132301 (2010).
%  [arXiv:1006.4783 [nucl-th]].
  %%CITATION = PRLTA,105,132301;%%
% 9
\bibitem{Juszczak:2010ve}
  C.~Juszczak, J.~T.~Sobczyk and J.~Zmuda,
  %``On extraction of value of axial mass from MiniBooNE neutrino quasi-elastic
  %double differential cross section data,''
  Phys.\ Rev.\  C {\bf 82}, 045502 (2010).
% [arXiv:1007.2195 [nucl-th]].
  %%CITATION = PHRVA,C82,045502;%%
% 10
\bibitem{Ankowski:2011dc}
  A.~M.~Ankowski and O.~Benhar,
  %``Electroweak nuclear response at moderate momentum transfer,''
  Phys.\ Rev.\  C {\bf 83}, 054616 (2011).
 % [arXiv:1102.3532 [nucl-th]].
  %%CITATION = PHRVA,C83,054616;%%
% 11
\bibitem{Alberico:1997vh}
  W.~M.~Alberico {\it et al.},
  %``Inelastic nu and anti-nu scattering on nuclei and *strangeness* of the
  %nucleon,''
  Nucl.\ Phys.\  A {\bf 623}, 471 (1997).
% 12
\bibitem{Alberico:1997rm}
  W.~M.~Alberico {\it et al.},
  %``The ratio of p and n yields in NC neutrino (anti-neutrino) nucleus
  %scattering and strange form-factors of the nucleon,''
  Phys.\ Lett.\  B {\bf 438}, 9 (1998).
  %%CITATION = PHLTA,B438,9;%%
% 13
\bibitem{Alberico:1998qw}
  W.~M.~Alberico {\it et al.},
  %``Strange form-factors of the proton: A New analysis of the neutrino
  %(anti-neutrino) data of the BNL-734 experiment,''
  Nucl.\ Phys.\  A {\bf 651}, 277 (1999).
  %%CITATION = NUPHA,A651,277;%%
% 14
\bibitem{Meucci:2003cv}
  A.~Meucci, C.~Giusti and F.~D.~Pacati,
  %``Relativistic Green's function approach to charged-current neutrino  nucleus
  %quasielastic scattering,''
  Nucl.\ Phys.\  A {\bf 739}, 277 (2004).
% 15
\bibitem{Meucci:2004ip}
  A.~Meucci, C.~Giusti and F.~D.~Pacati,
  %``Neutral-current neutrino-nucleus quasielastic scattering,''
  Nucl.\ Phys.\  A {\bf 744}, 307 (2004).
% 16
\bibitem{Meucci:2006ir}
  A.~Meucci, C.~Giusti and F.~D.~Pacati,
  %``Neutrino nucleus quasi-elastic scattering and strange quark effects,''
  Nucl.\ Phys.\  A {\bf 773}, 250 (2006).
% 17
\bibitem{Meucci:2011pi}
  A.~Meucci, J.~A.~Caballero, C.~Giusti and J.~M.~Udias,
  %``Relativistic descriptions of quasielastic charged-current neutrino-nucleus
  %scattering: application to scaling and superscaling ideas,''
  Phys.\ Rev.\  C {\bf 83}, 064614 (2011).
%  [arXiv:1103.0636 [nucl-th]].
  %%CITATION = PHRVA,C83,064614;%%
% 18
\bibitem{Meucci:2011vd}
  A.~Meucci, M.~B.~Barbaro, J.~A.~Caballero, C.~Giusti and J.~M.~Udias,
  %``Relativistic descriptions of final-state interactions in charged-current
  %quasielastic neutrino-nucleus scattering at MiniBooNE kinematics,''
  arXiv:1107.5145 [nucl-th].
  %%CITATION = ARXIV:1107.5145;%%
% 19
\bibitem{Antonov:2006md}
  A.~N.~Antonov {\it et al.},
  %``Superscaling analysis of inclusive electron scattering and its extension to
  %charge-changing neutrino-nucleus cross sections beyond the relativistic Fermi
  %gas approach,''
  Phys.\ Rev.\  C {\bf 74}, 054603 (2006).
  %%CITATION = PHRVA,C74,054603;%%
% 20
\bibitem{Antonov:2007vd}
  A.~N.~Antonov, M.~V.~Ivanov, M.~B.~Barbaro, J.~A.~Caballero, E.~Moya de Guerra and M.~K.~Gaidarov,
  %``Superscaling and Neutral Current Quasielastic Neutrino-Nucleus Scattering
  %beyond the Relativistic Fermi Gas Model,''
  Phys.\ Rev.\  C {\bf 75}, 064617 (2007).
  %%CITATION = PHRVA,C75,064617;%%
% 21
\bibitem{Ivanov:2008ng}
  M.~V.~Ivanov, M.~B.~Barbaro, J.~A.~Caballero, A.~N.~Antonov, E.~Moya de Guerra and M.~K.~Gaidarov,
  %``Superscaling and Charge-Changing Neutrino Scattering from Nuclei in the
  %Delta-Region beyond the Relativistic Fermi Gas Model,''
  Phys.\ Rev.\  C {\bf 77}, 034612 (2008).
  %%CITATION = PHRVA,C77,034612;%%
% 22
\bibitem{Leitner:2006ww}
  T.~Leitner, L.~Alvarez-Ruso and U.~Mosel,
  %``Charged current neutrino nucleus interactions at intermediate energies,''
  Phys.\ Rev.\  C {\bf 73}, 065502 (2006).
  %%CITATION = PHRVA,C73,065502;%%
% 23
\bibitem{Leitner:2006sp}
  T.~Leitner, L.~Alvarez-Ruso and U.~Mosel,
  %``Neutral current neutrino-nucleus interactions at intermediate energies,''
  Phys.\ Rev.\  C {\bf 74}, 065502 (2006).
  %%CITATION = PHRVA,C74,065502;%%
% 24
\bibitem{Buss:2007ar}
  O.~Buss, T.~Leitner, U.~Mosel and L.~Alvarez-Ruso,
  %``The Influence of the nuclear medium on inclusive electron and neutrino
  %scattering off nuclei,''
  Phys.\ Rev.\  C {\bf 76}, 035502 (2007).
  %%CITATION = PHRVA,C76,035502;%%
% 25
\bibitem{Leitner:2010kp}
  T.~Leitner and U.~Mosel,
  %``Neutrino-nucleus scattering reexamined: Quasielastic scattering and pion
  %production entanglement and implications for neutrino energy
  %reconstruction,''
  Phys.\ Rev.\  C {\bf 81}, 064614 (2010).
  %%CITATION = PHRVA,C81,064614;%%
% 26
\bibitem{Nieves:2004wx}
  J.~Nieves, J.~E.~Amaro and M.~Valverde,
  %``Inclusive quasi-elastic neutrino reactions,''
  Phys.\ Rev.\  C {\bf 70}, 055503 (2004)
  [Erratum-ibid.\  C {\bf 72}, 019902 (2005)].
  %%CITATION = PHRVA,C70,055503;%%
% 27
\bibitem{Nieves:2005rq}
  J.~Nieves, M.~Valverde and M.~J.~Vicente Vacas,
  %``Inclusive nucleon emission induced by quasi-elastic neutrino-nucleus
  %interactions,''
  Phys.\ Rev.\  C {\bf 73}, 025504 (2006).
  %%CITATION = PHRVA,C73,025504;%%
% 28
\bibitem{Valverde:2006zn}
  M.~Valverde, J.~E.~Amaro and J.~Nieves,
  %``Theoretical uncertainties on quasielastic charged-current neutrino-nucleus
  %cross sections,''
  Phys.\ Lett.\  B {\bf 638}, 325 (2006).
  %%CITATION = PHLTA,B638,325;%%
% 29
\bibitem{Martini:2009uj}
  M.~Martini, M.~Ericson, G.~Chanfray and J.~Marteau,
  %``A Unified approach for nucleon knock-out, coherent and incoherent pion
  %production in neutrino interactions with nuclei,''
  Phys.\ Rev.\  C {\bf 80}, 065501 (2009).
  %%CITATION = PHRVA,C80,065501;%%
% 30
\bibitem{Martini:2010ex}
  M.~Martini, M.~Ericson, G.~Chanfray and J.~Marteau,
  %``Neutrino and antineutrino quasielastic interactions with nuclei,''
  Phys.\ Rev.\  C {\bf 81}, 045502 (2010).
  %%CITATION = PHRVA,C81,045502;%%
% 31
\bibitem{Nieves:2011pp}
  J.~Nieves, I.~Ruiz Simo and M.~J.~Vicente Vacas,
  %``Inclusive Charged--Current Neutrino--Nucleus Reactions,''
  Phys.\ Rev.\  C {\bf 83}, 045501 (2011).
  %%CITATION = PHRVA,C83,045501;%%
% 32
\bibitem{Nieves:2011yp}
  J.~Nieves, I.~R.~Simo and M.~J.~V.~Vacas,
  %``The nucleon axial mass and the MiniBooNE Quasielastic Neutrino-Nucleus
  %Scattering problem,''
  arXiv:1106.5374 [hep-ph].
  %%CITATION = ARXIV:1106.5374;%%
% 33
\bibitem{Giusti:2011ar}
  C.~Giusti and A.~Meucci,
  %``Models for quasielastic electron and neutrino-nucleus scattering,''
  arXiv:1107.5425 [nucl-th].
  %%CITATION = ARXIV:1107.5425;%%
% 34
\bibitem{Amaro:1998ta}
  J.~E.~Amaro, M.~B.~Barbaro, J.~A.~Caballero, T.~W.~Donnelly and A.~Molinari,
  %``Relativistic effects in electromagnetic meson exchange currents,''
  Nucl.\ Phys.\  A {\bf 643}, 349 (1998).
% [arXiv:nucl-th/9806014].
  %%CITATION = NUPHA,A643,349;%%
% 35
\bibitem{Amaro:2002mj}
  J.~E.~Amaro, M.~B.~Barbaro, J.~A.~Caballero, T.~W.~Donnelly and A.~Molinari,
  %``Gauge and Lorentz invariant one-pion exchange currents in electron
  %scattering from a relativistic Fermi gas,''
  Phys.\ Rept.\  {\bf 368}, 317 (2002).
% 36
\bibitem{Day:1990mf}
  D.~B.~Day, J.~S.~McCarthy, T.~W.~Donnelly and I.~Sick,
  %``Scaling in inclusive electron - nucleus scattering,''
  Ann.\ Rev.\ Nucl.\ Part.\ Sci.\  {\bf 40}, 357 (1990).
% 37
\bibitem{Jourdan:1996ut}
  J.~Jourdan,
  %``Quasielastic response functions: The Coulomb sum revisited,''
  Nucl.\ Phys.\  A {\bf 603} (1996) 117.
% 38
\bibitem{Donnelly:1998xg}
  T.~W.~Donnelly and I.~Sick,
  %``Superscaling in inclusive electron nucleus scattering,''
  Phys.\ Rev.\ Lett.\  {\bf 82}, 3212 (1999).
% 39
\bibitem{Donnelly:1999sw}
  T.~W.~Donnelly and I.~Sick,
  %``Superscaling of inclusive electron scattering from nuclei,''
  Phys.\ Rev.\  C {\bf 60}, 065502 (1999).
  %%CITATION = PHRVA,C60,065502;%%
% 40
\bibitem{Maieron:2001it}
  C.~Maieron, T.~W.~Donnelly and I.~Sick,
  %``Extended superscaling of electron scattering from nuclei,''
  Phys.\ Rev.\  C {\bf 65} (2002) 025502.
  %%CITATION = PHRVA,C65,025502;%%
% 41
\bibitem{Alberico:1988bv}
  W.~M.~Alberico, A.~Molinari, T.~W.~Donnelly, E.~L.~Kronenberg and J.~W.~Van Orden,
  %``Scaling in electron scattering from a relativistic Fermi gas,''
  Phys.\ Rev.\  C {\bf 38}, 1801 (1988).
% 42
\bibitem{Barbaro:2006me}
  M.~B.~Barbaro, J.~E.~Amaro, J.~A.~Caballero, T.~W.~Donnelly,
  %``Superscaling in lepton-nucleus scattering,''
 Proceedings of the XXV International Worshop on Nuclear Theory, Rila, 2005,
 ``Nuclear Theory 25'', Ed. S.~Dimitrova, Heron Press, Sofia, 2006; 
 [arXiv:nucl-th/0609057].
  %%CITATION = NUCL-TH/0609057;%%
% 43
\bibitem{Amaro:2010sd}
  J.~E.~Amaro, M.~B.~Barbaro, J.~A.~Caballero, T.~W.~Donnelly and C.~F.~Williamson,
  %``Meson-exchange currents and quasielastic neutrino cross sections in the
  %SuperScaling Approximation model,''
  Phys.\ Lett.\  B {\bf 696}, 151 (2011).
%  [arXiv:1010.1708 [nucl-th]].
  %%CITATION = PHLTA,B696,151;%%
% 44
\bibitem{Amaro:2011qb}
  J.~E.~Amaro, M.~B.~Barbaro, J.~A.~Caballero, T.~W.~Donnelly and J.~M.~Udias,
  %``Relativistic analyses of quasielastic neutrino cross sections at MiniBooNE
  %kinematics,''
  Phys.\ Rev.\ D {\bf 84}, 033004 (2011).
%  arXiv:1104.5446 [nucl-th].
  %%CITATION = ARXIV:1104.5446;%%
% 45
\bibitem{Amaro:2004bs}
  J.~E.~Amaro, M.~B.~Barbaro, J.~A.~Caballero, T.~W.~Donnelly, A.~Molinari and I.~Sick,
  %``Using electron scattering superscaling to predict charge-changing  neutrino
  %cross sections in nuclei,''
  Phys.\ Rev.\  C {\bf 71} (2005) 015501.
  %%CITATION = PHRVA,C71,015501;%%
% 46
\bibitem{Caballero:2006wi}
  J.~A.~Caballero,
  %``General study of superscaling in quasielastic (e,e') and (nu, mu) reactions
  %using the relativistic impulse approximation,''
  Phys.\ Rev.\  C {\bf 74}, 015502 (2006).
 % [arXiv:nucl-th/0604020].
  %%CITATION = PHRVA,C74,015502;%%
% 47
\bibitem{Caballero:2007tz}
  J.~A.~Caballero, J.~E.~Amaro, M.~B.~Barbaro, T.~W.~Donnelly and J.~M.~Udias,
  %``Scaling and isospin effects in quasielastic lepton-nucleus scattering in
  %the Relativistic Mean Field Approach,''
  Phys.\ Lett.\  B {\bf 653}, 366 (2007).
%  [arXiv:0705.1429 [nucl-th]].
  %%CITATION = PHLTA,B653,366;%%
% 48
\bibitem{Caballero:2005sj}
  J.~A.~Caballero, J.~E.~Amaro, M.~B.~Barbaro, T.~W.~Donnelly, C.~Maieron and J.~M.~Udias,
  %``Superscaling in charged current neutrino quasielastic scattering in the
  %relativistic impulse approximation,''
  Phys.\ Rev.\ Lett.\  {\bf 95}, 252502 (2005).
% 49
\bibitem{Amaro:2006tf}
  J.~E.~Amaro, M.~B.~Barbaro, J.~A.~Caballero and T.~W.~Donnelly,
  %``Quasielastic charged current neutrino nucleus scattering,''
  Phys.\ Rev.\ Lett.\  {\bf 98} (2007) 242501.
% 50
\bibitem{Maieron:2003df}
  C.~Maieron, M.~C.~Martinez, J.~A.~Caballero and J.~M.~Udias,
  %``Nuclear model effects in charged current neutrino nucleus quasielastic
  %scattering,''
  Phys.\ Rev.\  C {\bf 68}, 048501 (2003).
  %%CITATION = PHRVA,C68,048501;%%
% 51
\bibitem{Amaro:2006if}
  J.~E.~Amaro, M.~B.~Barbaro, J.~A.~Caballero, T.~W.~Donnelly and J.~M.~Udias,
  %``Final-state interactions and superscaling in the semi-relativistic approach
  %to quasielastic electron and neutrino scattering,''
  Phys.\ Rev.\  C {\bf 75}, 034613 (2007).
% 52
\bibitem{Horowitz}
C.~J.~Horowitz, B.~D.~Serot, Nucl. Phys. A {\bf 368}, 503 (1981); Phys. Lett. B {\bf 86}, 146 (1979).
% 53
\bibitem{Serot}
B.~D.~Serot, J.~D.~Walecka, Adv. Nucl. Phys. {\bf 16}, 1.
Eds. J.~W.~Negele, E.~W.~Vogt. Plenum Press, New York (1986).
% 54
\bibitem{Sharma:1993it}
  M.~M.~Sharma, M.~A.~Nagarajan and P.~Ring,
  %``rho meson coupling in the relativistic mean field theory and description of
  %exotic nuclei,''
  Phys.\ Lett.\  B {\bf 312}, 377 (1993).
  %%CITATION = PHLTA,B312,377;%%
% 55
\bibitem{Amaro:2006pr}
  J.~E.~Amaro, M.~B.~Barbaro, J.~A.~Caballero and T.~W.~Donnelly,
  %``Superscaling and neutral current quasielastic neutrino nucleus
  %scattering,''
  Phys.\ Rev.\  C {\bf 73}, 035503 (2006).
% 56
\bibitem{Amaro:2005dn}
  J.~E.~Amaro, M.~B.~Barbaro, J.~A.~Caballero, T.~W.~Donnelly and C.~Maieron,
  %``Semi-relativistic description of quasielastic neutrino reactions and
  %superscaling in a continuum shell model,''
  Phys.\ Rev.\  C {\bf 71}, 065501 (2005).
% 57
\bibitem{Amaro:2010iu}
  J.~E.~Amaro, C.~Maieron, M.~B.~Barbaro, J.~A.~Caballero and T.~W.~Donnelly,
  %``Pionic correlations and meson-exchange currents in two-particle emission
  %induced by electron scattering,''
  Phys.\ Rev.\  C {\bf 82}, 044601 (2010).
%  [arXiv:1008.0753 [nucl-th]].
  %%CITATION = PHRVA,C82,044601;%%
% 58
\bibitem{Alberico:1989aja}
  W.~M.~Alberico, T.~W.~Donnelly and A.~Molinari,
  %``PIONIC EFFECTS IN QUASIELASTIC ELECTRON SCATTERING,''
  Nucl.\ Phys.\  A {\bf 512}, 541 (1990).
  %CITATION = NUPHA,A512,541;%%
% 59
\bibitem{Amaro:2003yd}
  J.~E.~Amaro, M.~B.~Barbaro, J.~A.~Caballero, T.~W.~Donnelly and A.~Molinari,
  %``Delta-isobar relativistic meson-exchange currents in quasielastic  electron
  %scattering,''
  Nucl.\ Phys.\  A {\bf 723}, 181 (2003).
% 60
\bibitem{Amaro:2009dd}
  J.~E.~Amaro, M.~B.~Barbaro, J.~A.~Caballero, T.~W.~Donnelly, C.~Maieron and J.~M.~Udias,
  %``Meson-exchange currents and final-state interactions in quasielastic
  %electron scattering at high momentum transfers,''
  Phys.\ Rev.\  C {\bf 81} (2010) 014606.
%  [arXiv:0906.5598 [nucl-th]].
% 61
\bibitem{Donnelly:1978xa}
  T.~W.~Donnelly, J.~W.~Van Orden, T.~.~J.~De Forest and W.~C.~Hermans,
  %``MESON EXCHANGE CURRENTS IN DEEP INELASTIC ELECTRON SCATTERING FROM
  %NUCLEI,''
  Phys.\ Lett.\  B {\bf 76}, 393 (1978).
  %%CITATION = PHLTA,B76,393;%%
% 62
\bibitem{VanOrden:1980tg}
  J.~W.~Van Orden and T.~W.~Donnelly,
  %``MESONIC PROCESSES IN DEEP INELASTIC ELECTRON SCATTERING FROM NUCLEI,''
  Annals Phys.\  {\bf 131} (1981) 451.
  %%CITATION = APNYA,131,451;%%
% 63
\bibitem{Dekker:1994yc}
  M.~J.~Dekker, P.~J.~Brussaard and J.~A.~Tjon,
  %``Relativistic meson exchange and isobar currents in electron scattering:
  %Noninteracting Fermi gas analysis,''
  Phys.\ Rev.\  C {\bf 49}, 2650 (1994).
% 64
\bibitem{De Pace:2003xu}
  A.~De Pace, M.~Nardi, W.~M.~Alberico, T.~W.~Donnelly and A.~Molinari,
  %``The 2p-2h electromagnetic response in the quasielastic peak and beyond,''
  Nucl.\ Phys.\  A {\bf 726}, 303 (2003).
  %%CITATION = NUPHA,A726,303;%%
% 65
\bibitem{De Pace:2004cr}
  A.~De Pace, M.~Nardi, W.~M.~Alberico, T.~W.~Donnelly and A.~Molinari,
  %``Role of 2p-2h MEC excitations in superscaling,''
  Nucl.\ Phys.\  A {\bf 741}, 249 (2004).
  %%CITATION = NUPHA,A741,249;%%
% 66
\bibitem{Alberico:1983zg}
  W.~M.~Alberico, M.~Ericson and A.~Molinari,
  %``THE ROLE OF TWO PARTICLES - TWO HOLES EXCITATIONS IN THE SPIN - ISOSPIN
  %NUCLEAR RESPONSE,''
  Annals Phys.\  {\bf 154}, 356 (1984).
  %%CITATION = APNYA,154,356;%%
% 67
\bibitem{Alberico:1990fc}
  W.~M.~Alberico, A.~De Pace, A.~Drago and A.~Molinari,
  %``SECOND ORDER EFFECTS IN THE NUCLEAR RESPONSE FUNCTIONS,''
  Riv.\ Nuovo Cim.\  {\bf 14N5}, 1 (1991).
  %%CITATION = RNCIB,14N5,1;%%
% 68
\bibitem{Gil:1997bm}
  A.~Gil, J.~Nieves and E.~Oset,
  %``Many body approach to the inclusive (e, e-prime) reaction from the
  %quasielastic to the Delta excitation region,''
  Nucl.\ Phys.\  A {\bf 627}, 543 (1997).
%  [arXiv:nucl-th/9711009].
  %%CITATION = NUPHA,A627,543;%%
% 69
\bibitem{Amaro:wip}
  J.~E.~Amaro et al., work in progress.
% 70
\bibitem{Bodek:2011ps}
  A.~Bodek, H.~Budd, E.~Christy,
  %``Neutrino Quasielastic Scattering on Nuclear Targets: Parametrizing
  %Transverse Enhancement (Meson Exchange Currents),''
  arXiv:1106.0340 [hep-ph].
  %%CITATION = ARXIV:1106.0340;%%
\end{thebibliography}
\end{document}